\documentstyle[sprocl,psfig]{article}

\def\be{\begin{equation}}
\def\ee{\end{equation}}
\def\bea{\begin{eqnarray}}
\def\eea{\end{eqnarray}}

\begin{document}

\title{IMPROVED LATTICE QCD ACTIONS FOR HADRON PHENOMENOLOGY}

\author{DEREK B. LEINWEBER}

\address{Department of Physics and Math.\ Physics,
University of Adelaide 5005,
Australia\\
E-mail:~dleinweb@physics.adelaide.edu.au} 

\author{FRANK X. LEE}

\address{Nuclear Physics Laboratory, Department of Physics,
University of Colorado, \\Boulder, CO
80309-0446\\
E-mail:~fxlee@sammy.colorado.edu} 


\maketitle 

\vspace{-6.0cm}
\hfill ADP-98-53/T321
\vspace{5.5cm}

\abstracts{ The masses and dispersions of light hadrons are
calculated in lattice QCD using an $O(a^2)$ tadpole-improved gluon
action and the D$\chi34$ action, an $O(a^2)$ tadpole-improved
next-nearest-neighbor fermion action originally proposed by Hamber and
Wu.  Two lattices of constant volume with lattice spacings of
approximately 0.40 fm and 0.24 fm are considered.  The results reveal
some scaling violations at the coarser lattice spacing on the order of
5\%.  At the finer lattice spacing, the calculated $N/\rho$ mass ratio
reproduces state-of-the-art results using unimproved actions.  Good
dispersion and rotational invariance up to momenta of $pa \simeq 1$
are also found.  The relative merit of alternative choices for
improvement operators is assessed through close comparisons with other
plaquette-based tadpole-improved actions.  
}

\section{Introduction}

Lattice discretization of the continuum QCD action introduces errors
at finite lattice spacing $a$.  The standard Wilson gauge action has
$O(a^2)$ errors and the standard Wilson fermion action has $O(a)$
errors.  Simulations using these actions have shown that lattice
spacings of 0.1 fm or less and lattice volumes of $24^4$ or larger are
needed in order to hold systematic errors to the 10\% level.  Such
simulations are major undertakings \cite{Butler94,Yos97} and require
enormous computing power to extract even the most basic of hadronic
observables, the hadron masses.  
\footnotetext{ Presented by D.B.L. at
the workshop on ``Nonperturbative Methods in Quantum Field Theory,''
CSSM, Adelaide, Feb.\ 2--13, 1998.  This and related papers may be
obtained from:

\noindent
{\sf
http://www.physics.adelaide.edu.au/theory/staff/leinweber/publications.html}
}

During the past few years, considerable efforts have been devoted to
improving lattice actions \cite{Lep97,Has97}.  The idea is to reduce or
remove the discretization errors from the actions so that they have
better continuum-like behavior.  At the same time, errors due to the
lattice regularization are accounted for through the renormalization
of the coefficients multiplying the improvement operators.  The hope
is the use of improved actions will allow one to simulate efficiently
and accurately on coarse lattices, such that computer resources may be
redirected to the simulation of QCD rather than quenched QCD.
Moreover, one may turn the focus of investigation towards quantities
of experimental interest.

In the pure gauge sector, the $O(a^2)$ tadpole-improved
action \cite{Alf95} leads to dramatic improvement in the static
potential and glueball masses \cite{Mor97} up to lattice spacings of
0.4 fm.  In the light quark sector, hadron spectroscopy has been
investigated with a variety of improved actions including the
$O(a)$-improved SW action \cite{Col97}, the $O(a^2)$-improved D234
action \cite{Alf95a} and its variants \cite{Lep97,Alf96}, and the
D$\chi34$ action of Hamber and Wu \cite{Ham83,Egu84,Fie96,Lee97}
considered here.

   The D$\chi34$ action is an $O(a^2)$ next-nearest-neighbor fermion
action with tadpole-improved estimates of the coupling
renormalizations.  This action is selected primarily due to its
simplicity.  The cost of simulating it is about a factor of two as
compared to standard Wilson fermions.  Our goal is to study its
feasibility as an alternative action to SW which has the clover term,
or to D234 which has both next-nearest-neighbor couplings and the
clover term.  In particular, we examine dispersion relations and test
the rotational symmetry of both the gauge and the fermion actions.
Hadron mass ratios are calculated for a wide variety of hadrons
including hyperons.  To explore scaling violations, we consider two
coarse lattices of approximately fixed physical volume: $6^3\times 12$
at a lattice spacing of 0.40 fm and $10^3\times 16$ at 0.24 fm.

\section{Improved Lattice Actions}

The improved gauge action employed in this investigation is given
by \cite{Alf95}:
\begin{equation}
S_G=\beta\sum_{\mbox{pl}}{1\over 3} \mbox{Re Tr}(1-U_{\mbox{pl}})
-{\beta\over 20 u^2_0}\sum_{\mbox{rt}}{1\over 3} 
\mbox{Re Tr}(1-U_{\mbox{rt}}) \, .
\end{equation}
The second term removes the $O(a^2)$ errors at tree level.
Perturbative corrections are estimated \cite{Alf95} to be of the order
of 2-3\%.  Here, $U_{\mbox{rt}}$ denotes the rectangular 1x2
plaquettes.  $u_0$ is the tadpole factor that largely corrects for the
large quantum renormalization of the links $U_\mu(x)=\exp(i \, g\,
\int_x^{x+a \widehat\mu} A(y) \cdot dy )$.  In this calculation we use
the mean plaquette $u_0 \equiv ( \mbox{Re Tr} \langle
U_{\mbox{pl}}\rangle )^{1/4}/3$ to estimate $u_0$, and will focus our
evaluation of lattice action improvement on other plaquette-based
improved actions.  $u_0$ is determined self-consistently in the
simulation.

The improvement program of Sheikoleslami and Wohlert (SW) \cite{She85}
provides a systematic approach to the improvement of lattice fermion
actions.  However, the on-shell improvement program leaves some
freedom in the relative values of the coefficients of the improvement
operators.  In this investigation, we consider a specific case of the
general class of D234 actions \cite{Alf96} in which the improvement
parameters are tuned to remove the second-order
chiral-symmetry-breaking Wilson term at tree level.  This fermion
action may be written
\begin{equation}
M_{\rm{D}\chi34} = m_q + \gamma \cdot \nabla + {1 \over 6} \sum_\mu
\left ( -a^2 \nabla_\mu \Delta_\mu + b \, a^3 \Delta_\mu^2 \right ) \,
\label{Dchi34}
\end{equation}
where
\begin{equation}
\nabla_\mu \, \psi(x) = {1 \over 2a \, u_0} \left [
U_\mu(x) \, \psi(x+\mu) - U_\mu^\dagger(x-\mu) \, \psi(x-\mu) \right ]
\, ,
\end{equation}
and
\begin{equation}
\Delta_\mu \, \psi(x) = {1 \over a^2\, u_0} \left [
U_\mu(x) \, \psi(x+\mu) + U_\mu^\dagger(x-\mu) \, \psi(x-\mu) - 2 \,
u_0\, \psi(x) \right ] \, .
\end{equation}
The second-order term of the D234 action, $\sum_\mu \Delta_\mu +
\sigma \cdot F /2$, breaks chiral symmetry and does not appear in the
D$\chi34$ action.  The fourth-order term of (\ref{Dchi34}) breaks
chiral symmetry and provides for the removal of the fermion doublers.
The D$\chi34$ action is free of both $O(a)$ and $O(a^2)$ errors at
tree level.  Explicit evaluation of (\ref{Dchi34}) combined with a
Wilson-fermion-style field renormalization, discloses the fermion
action of Hamber and Wu \cite{Ham83,Egu84,Lee97}.

\begin{figure}[t]
\centerline{\psfig{file=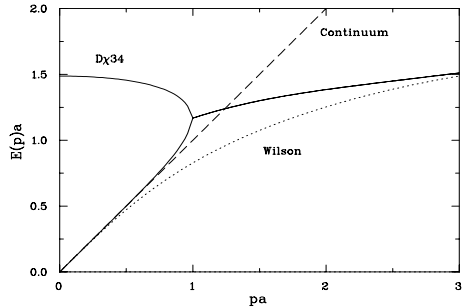,height=4cm,width=6cm,angle=90}}
\vspace{-0.1cm}
\caption{Free dispersion relations for zero mass D$\chi34$, Wilson, and
continuum fermions.  Momentum $p$ is along the (1,1,0) direction.
Beyond the D$\chi34$ branch point, the real part of the two conjugate
roots is shown.}
\label{disp}
\end{figure}

Free dispersion relations can be obtained by locating the poles in the
fermion propagator.  Fig.~\ref{disp} shows free dispersion relations
for massless continuum, Wilson, and D$\chi34$ fermions.  It is clear
that D$\chi34$ fermions follow the continuum more closely than Wilson
fermions.  Note that there exists an unphysical high energy doubler
(or ghost) in the D$\chi34$ action.  It is very similar to that for
the D234 action \cite{Alf95a} and is a general feature of fermions
with next-nearest-neighbor couplings.  The doubler can be `pushed
away' from the low momentum region by various techniques, such as
tuning the value of $b$, or using an anisotropic lattice \cite{Alf96}.
We simulate with $b = 1$, which gives good dispersion to $p a \sim
1$ as illustrated in Fig.~\ref{disp}.

\section{Lattice Simulations}

\subsection{Methods and Parameters}

Quenched gauge configurations are generated using the
Cabibbo-Marinari pseudo-heat-bath method \cite{Cab82}.  Periodic
boundary conditions are used in all directions for the gauge field
and in spatial directions for the fermion field.  Dirichlet boundary
conditions are used for the fermion field in the time direction.
Configurations separated by 300 sweeps are selected after 4000
thermalization sweeps from a cold start.

\begin{figure}[t]
\centerline{\psfig{file=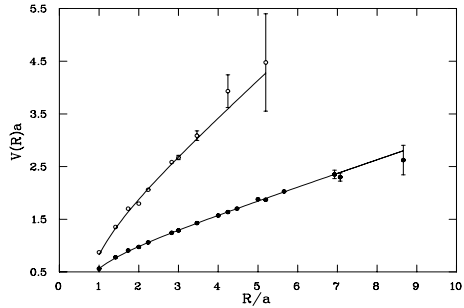,height=4cm,width=6cm,angle=90}}
\vspace{-0.1cm}
\caption{Static potential from Wilson loops.  The empty circles are
for $\beta=6.25$, the solid circles for $\beta=7.0$.  The lines are
best fits.  The statistical errors are from 200 configurations in
both cases.}
\label{pot}
\end{figure}

The static potential is calculated and the string tension $\sigma$ is
extracted from the ansatz $V(R) = V_0 + \sigma R - E/R$ where $V_0$
and $E$ are constants.  Fig.~\ref{pot} shows our results.  Good
rotational invariance of the static potential is observed.  Using
$\sqrt{\sigma}=440$ MeV to set the scale, the lattice spacings for the
coarse and fine lattices are 0.40(6) fm and 0.23(1) fm with
$\chi^2/N_{\rm DF}$ of 1.46 and 0.91 respectively.  We analyze 155
configurations on our coarse $6^3 \times 12$ lattice and 100
configurations on our fine $10^3 \times 16$ lattice.

Five quark propagators are computed by the Stabilized Biconjugate
Gradient algorithm \cite{From94} for each configuration.  The five
quark masses selected are approximately 210, 180, 150, 120, 90 MeV,
for both lattices.  The second value, 180 MeV, is taken as the strange
quark mass.  A point source is used at space-time location
(x,y,z,t)=(1,1,1,2) on the $6^3\times 12$ lattice and (1,1,1,3) on the
$10^3\times 16$ lattice.  The gauge-invariant smearing
method \cite{Gus90} is applied at the sink to increase the overlap of
the interpolating operators with the ground states.

Statistical errors are estimated in a third-order, single-elimination
jackknife, with bias corrections \cite{Jack}.  A third-order jackknife
provides uncertainty estimates for the correlation functions, fits to
the correlation functions, and quantities extrapolated to the chiral
limit.

\subsection{Hadron Masses}

\begin{figure}[t]
\centerline{\psfig{file=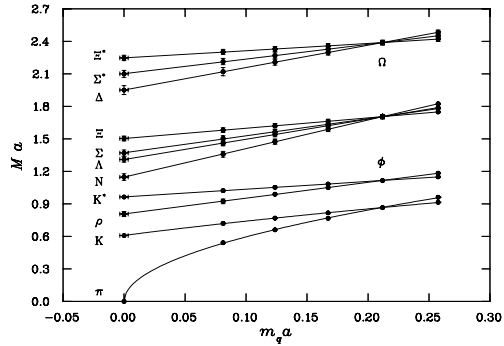,height=4.5cm,width=6.5cm,angle=90}}
\vspace{-0.1cm}
\caption{Hadron masses in lattice units as a function of $m_q$ on our
fine lattice.  The lines are chiral fits as discussed in the text. For
better viewing, the decuplet masses are shifted upward by 0.5 unit.}
\label{mall}
\end{figure}

Fig.~\ref{mall} shows the extracted hadron masses plotted as a
function of the quark mass.  We report values taken from covariance
matrix fits to the time slice interval 4 through 8 on our coarse
lattice and 6 through 9 on our fine lattice.  These regimes provide
the best signal-to-noise and good correlated $\chi^2/N_{\rm DF}$.
$\kappa_{cr}$ is determined by linearly extrapolating $m_\pi^2$ as a
function of $m_q$ to zero.  Similarly, the pseudoscalar kaon is
extrapolated via $m_K^2 = c_0 + c_1 \, m_q$.  The form $M=c_0+c_1 \,
m_q$ is used for all other extrapolations to the chiral limit.  Fits
including an additional term $c_2 m^{3/2}_q$ are also considered and
similar results are found with slightly larger error bars.  

Ratios of the chirally extrapolated masses are given in
Table~\ref{ratio} along with the ratios as observed in
nature \cite{PDG94}.  At $\beta=7.0$ the lattice spacing estimates
follow the familiar pattern having the value based on the string
tension lying between that of the $\rho$ and nucleon based values.
This is most likely an artifact of the quenched approximation.
However, at $\beta=6.25$ we find significant disagreement among the
values and an unusual reordering of values.

\begin{table}[t]
\caption{Mass ratios after extrapolation to the chiral limit.
$a_\rho$ and $a_N$ are lattice spacings in fm set by the rho mass (770
MeV) and the nucleon mass (938 MeV).}
\label{ratio}
\vspace{0.2cm}
\begin{center}
\footnotesize
\begin{tabular}{|llll|llll|}
\hline
          & \raisebox{0pt}[13pt][7pt]{$\beta$=6.25} & $\beta$=7.0  & Expt.&
          & \raisebox{0pt}[13pt][7pt]{$\beta$=6.25} & $\beta$=7.0  & Expt.\\
\hline
\multicolumn{4}{|l|}{Vector/Vector} 
&\multicolumn{4}{|l|}{\raisebox{0pt}[13pt][7pt]{Pseudoscalar/Vector}}\\
$K^*/\rho$      & 1.20(2)  & 1.20(2)  & 1.16 &$K/\rho$        & 0.76(2)  & 0.75(2)  & 0.64 \\
$\phi/\rho$     & 1.40(3)  & 1.39(3)  & 1.32 &                &&&\\[4pt]
\multicolumn{4}{|l|}{Octet/Octet} &\multicolumn{4}{|l|}{Octet/Vector}\\
$\Lambda/N$     & 1.16(2)  & 1.16(1)  & 1.19 &$N/\rho$        & 1.48(6)  & 1.44(5)  & 1.22 \\
$\Sigma/N$      & 1.19(2)  & 1.20(1)  & 1.27 &$\Lambda/\rho$  & 1.72(6)  & 1.66(5)  & 1.45 \\
$\Xi/N$         & 1.34(3)  & 1.33(2)  & 1.40 &$\Sigma/\rho$   & 1.76(6)  & 1.72(5)  & 1.55 \\
                &          &          &      &$\Xi/\rho$      & 1.97(6)  & 1.91(6)  & 1.71 \\[4pt]
\multicolumn{4}{|l|}{Decuplet/Decuplet} &\multicolumn{4}{|l|}{Decuplet/Vector}\\
$\Sigma^*/\Delta$ & 1.12(1)  & 1.11(1)  & 1.12 &$\Delta/\rho$   & 1.79(7)  & 1.77(5)  & 1.60 \\
$\Xi^*/\Delta$    & 1.24(3)  & 1.22(2)  & 1.24 &$\Sigma^*/\rho$ & 2.00(6)  & 1.97(6)  & 1.80 \\
$\Omega/\Delta$   & 1.34(3)  & 1.33(2)  & 1.36 &$\Xi^*/\rho$    & 2.20(7)  & 2.16(6)  & 1.99 \\
                  &          &          &      &$\Omega/\rho$   & 2.40(7)  & 2.35(6)  & 2.17 \\[4pt]
\multicolumn{4}{|l|}{String Tensions}          &\multicolumn{4}{|l|}{Decuplet/Octet}\\
$a_{st}$  &  0.40(3)     & 0.220(2)     &      &$\Delta/N$      & 1.21(5)  & 1.23(2)  & 1.31 \\
$a_\rho$  &  0.30(1)     & 0.205(6)     &      &$\Sigma^*/N$    & 1.35(5)  & 1.37(2)  & 1.47 \\
$a_N$     &  0.36(1)     & 0.242(6)     &      &$\Xi^*/N$       & 1.49(4)  & 1.50(3)  & 1.63 \\
                  &          &          &      &$\Omega/N$      & 1.63(4)  & 1.64(3)  & 1.78 \\[4pt]
\hline
\end{tabular}
\end{center}
\end{table}

Focusing first on ratios of hadrons having the same angular momentum,
we see very little change in the values as the lattice spacing is
decreased.  These ratios are also remarkably similar to those observed
in nature, despite the fact that these are quenched QCD calculations.
In addition, these ratios support our selection for the strange quark
mass.

This close resemblance to nature is not shared by ratios of hadrons
with different angular momentum.  All four classes of ratios
significantly disagree with those of nature.  Once again we see the
familiar quenched artifact of the Octet/Vector ratio being too large
and the Decuplet/Octet ratio being too small.  

\begin{figure}[tb]
\centerline{\psfig{file=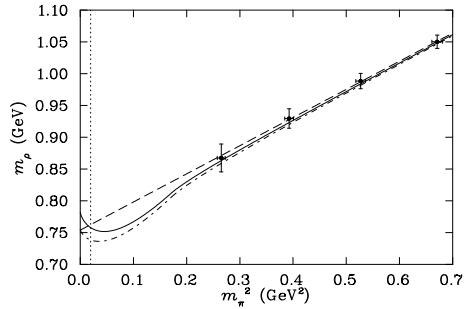,height=4cm,width=6cm,angle=90}}
\vspace{-0.1cm}
\caption{Plot of the $\rho$-meson mass as a function of the squared
pion mass obtained from our finer lattice.  $a_\rho$ has been used to
set the scale.  The dashed line illustrates the standard linear
extrapolations of $m_\pi^2$ and $m_\rho$.  The solid and dot-dash
curves include the two-pion self-energy of the $\rho$
meson~\protect\cite{Lein94} for dipole dispersion cut-off values of 1
and 2 GeV respectively.  The increase in the slope at $m_\rho/m_\pi =
1.8$ ($m_\pi^2 \simeq 0.21$) provided by the two-pion self energy is
the right order of magnitude to restore agreement with the empirical
value.}
\label{Jgraph}
\end{figure}

The standard failure of the $K/\rho$ mass ratio in the quenched
approximation is also seen here.  This shortcoming has been widely
realized through an examination of the $J$-parameter \cite{Lac95}
defined by
\begin{equation}
J = m_\rho \, {dm_\rho \over dm_\pi^2} \Bigm |_{m_\rho/m_\pi = 1.8} 
\;\simeq\; m_{K^*} \, {m_{K^*} - m_\rho \over m_{K}^2 - m_\pi^2 } \, .
\end{equation}
Empirically this ratio is 0.48.  However we find 0.42 on our coarse
lattice and 0.43 on our fine lattice.  The physics associated with
this discrepancy was first reported by Cohen and leinweber
\cite{Lein94} where it was pointed out that the self-energy generated
by two-pion intermediate states of the $\rho$-meson, which is excluded
in the quenched approximation, acts to increase the $J$ parameter.
Fig.~\ref{Jgraph} provides a sketch of how including the two-pion
self-energy of the $\rho$ can increase the value of $J$ to 0.46.

Perhaps the most important information displayed in Table~\ref{ratio}
is that the Octet/Vector mass ratios display less than satisfactory
scaling for the larger lattice spacing.  To further examine scaling
and make contact with other studies, we focus on the the $N/\rho$ mass
ratio which is among the the most revealing of ratios.

\begin{figure}[t]
\centerline{\psfig{file=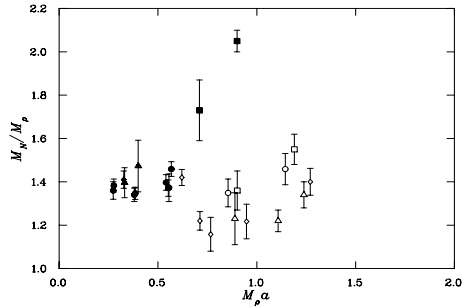,height=4cm,width=6cm,angle=90}}
\vspace{-0.1cm}
\caption{The $N/\rho$ mass ratio versus $M_\rho a$ at the chiral
limit.  Solid symbols denote the standard Wilson action.  Open symbols
denote improved actions including SW~\protect\cite{Col97} $(\diamond)$,
D234~\protect\cite{Alf95a} $(\triangle)$, D$\chi34$~\protect\cite{Fie96}
$(\Box)$, and D$\chi34$ $(\circ)$ (this work).}
\label{nrho}
\end{figure}

\begin{figure}[t]
\centerline{\psfig{file=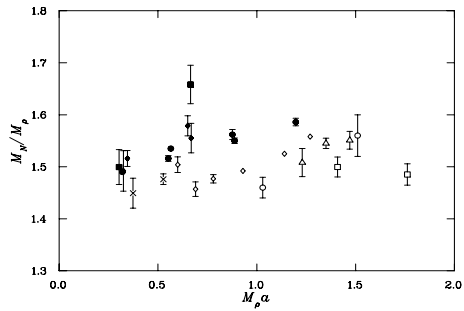,height=4cm,width=6cm,angle=90}}
\vspace{-0.1cm}
\caption{The $N/\rho$ mass ratio versus $M_\rho a$ at a fixed
$\pi/\rho$ mass ratio of 0.7 for various actions.  The solid symbols
denote the standard actions: Wilson (square and diamond), staggered
(circle).  The open symbols denote improved actions:
nonperturbatively-improved SW~\protect\cite{Goc97} $(\times)$,
fixed-point action~\protect\cite{DeG97} $(\Box)$, SW $(\diamond)$,
D234 $(\triangle)$, and D$\chi34$ $(\circ)$ (this work). }
\label{nrho07}
\end{figure}

Fig.~\ref{nrho} shows a comparison of the $N/\rho$ mass ratio versus
$M_\rho \, a$ at the chiral limit.  Fig.~\ref{nrho07} shows the
$N/\rho$ mass ratio as a function of $M_\rho \, a$ at a fixed
$\pi/\rho$ mass ratio \cite{DeG97} of 0.7.  This method is free of
complications from chiral extrapolations.  Both cases clearly show the
improvement provided by the D$\chi34$ action.  Indeed, the D$\chi34$
action has reproduced the state-of-the-art quenched QCD ratios using
unimproved actions at coarse lattice spacings of 0.24 fm.

\subsection{Dispersion and Rotational Symmetry}

In addition to mass ratios, hadron states at finite momentum
projections $\vec{p} \, a =\vec{n}(2\pi/L)$ are also calculated.
Dispersion is examined by calculating the effective speed of light,
defined by $c^2=(E^2(p)-E^2(0))/p^2$, which is to be compared with 1.

A comparison with SW and D234 lattice actions \cite{Alf95a} is made in
Table \ref{dispComp}.  The dispersion for the $O(a)$-improved SW
action is very poor relative to the excellent dispersions of the
next-nearest-neighbor improved D$x34$ actions in general.  The
D$\chi34$ dispersion is excellent even at our coarse lattice spacing.

\begin{table}[tb]
\caption{Comparison of SW, D234 and D$\chi34$ actions for the speed of
light squared obtained from the dispersion of $\pi$ and $\rho$ mesons
at $m_\pi/m_\rho \simeq 0.7$ for $p \, a=(2\pi/L)$.}
\label{dispComp}
\vspace{0.2cm}
\begin{center}
\footnotesize
\begin{tabular}{|ccccc|}
\hline
\raisebox{0pt}[13pt][7pt]{
Hadron} &$a$ (fm) &SW         &D234         &D$\chi34$  \\
\hline
$\pi$   &0.40     &0.63(2)    &0.95(2)      &\raisebox{0pt}[12pt][6pt]{
0.99(3)}    \\
$\pi$   &0.24     &           &0.99(4)      &1.04(4)    \\[4pt]
\hline
$\rho$  &0.40     &0.48(4)    &0.93(3)      &\raisebox{0pt}[12pt][6pt]{
0.93(6)}    \\
$\rho$  &0.24     &           &1.00(6)      &0.99(6)    \\[4pt]
\hline
\end{tabular}
\end{center}
\end{table}

Rotational symmetry is explored in Table~\ref{dispt}.  At the coarser
lattice spacing, some drift in the central values is seen for the pion
and nucleon.  The drift in the pion is similar to that seen for the
D234c action \cite{Lep97}.  However, the drift in dispersion previously
reported for the $\phi$ meson \cite{Lep97} is not apparent in our
results for the D$\chi34$ action.  The D$\chi34$ action has much
better rotational symmetry than the SW action \cite{Lep97}.  The
D$\chi34$ action provides satisfactory dispersion at our finer lattice
spacing and is competitive with the D234 action \cite{Alf95a}.

\begin{table}[t]
\caption{Evaluation of dispersion and rotational invariance via the
effective speed of light.  The results are for $m_q\sim$ 180 MeV.}
\label{dispt}
\vspace{0.2cm}
\begin{center}
\footnotesize
\begin{tabular}{|cccccc|}
\hline
\raisebox{0pt}[13pt][7pt]{
$a$ (fm)} & $\vec{n}$& $\pi$    & $\phi$   & $N$     & $\Omega$  \\
\hline
\raisebox{0pt}[12pt][6pt]{
0.40}    & (1,0,0)  & 0.98(2)  & 0.91(4)  & 1.00(7)  & 0.99(12) \\
	 & (1,1,0)  & 0.91(4)  & 0.91(6)  & 0.94(8)  & 0.91(7)  \\
	 & (1,1,1)  & 0.86(9)  & 0.92(10) & 0.92(7)  & 0.90(10) \\[4pt]
\hline
\raisebox{0pt}[12pt][6pt]{
0.24}    & (1,0,0)  & 1.04(3)  & 1.02(4)  & 1.10(6)  & 1.06(7) \\
	 & (1,1,0)  & 1.05(4)  & 1.02(4)  & 1.06(5)  & 1.11(5) \\
	 & (1,1,1)  & 0.98(6)  & 0.98(6)  & 1.06(6)  & 1.06(5) \\[4pt]
\hline
\end{tabular}
\end{center}
\end{table}

\section{Conclusion}

We have computed masses and dispersion relations of light hadrons in
lattice QCD using tree-level $O(a^2)$ tadpole-improved gauge and
fermion actions.  These actions have the appeal of being simple to
implement and inexpensive to simulate.  A great deal of effort is
being directed toward finding the ultimate improved action that will
facilitate simulations on the coarsest of lattices.  We note however,
that many quantities of phenomenological interest such as hadron form
factors involve momenta on the order of a GeV.  As such, a highly
improved action which is costly to simulate may not be the ideal
action for hadron phenomenology, especially for exploratory purposes.

The mass ratios obtained from the D$\chi34$ action at 0.24 fm on a
modest $10^3 \times 16$ lattice reproduce the state-of-the-art results
using conventional unimproved actions.  Excellent dispersion and
rotational invariance up to $pa\approx 1$ are also found.  These
results demonstrate that the D$\chi34$ action can serve as a viable 
candidate for the study of hadron phenomenology, 
and in our view is preferable to 
the highly-improved but more costly D234 action.  
We plan to use the D$\chi34$ action to study
hadron properties beyond the spectrum, such as multipole form factors
of hadrons in general.  These results also bode well for future
explorations beyond the quenched approximation.

\medskip
Support from the U.S. DOE and the Australian
Research Council is gratefully acknowledged.

\end{document}